\documentclass[final,5p,times,authoryear,12pt,twocolumn]{elsarticle}

\usepackage{natbib}
\usepackage{graphicx}
\usepackage[ansinew]{inputenc}
\usepackage{hyperref}
\usepackage{booktabs}
\usepackage{amsmath}
\usepackage{color}
\def\WmK{$\rm W\,m^{-1}\,K^{-1}$}

\journal{Icarus}

\begin{document}

\begin{frontmatter}



\title{Thermal conductivity measurements of porous dust aggregates: \\I. Technique, model and first results}


\author[igep]{M. Krause}
\ead{maya.krause@tu-bs.de}
\author[igep]{J. Blum}
\author[igep]{Yu. V. Skorov}
\author[heid]{M. Trieloff}
\address[igep]{Institut für Geophysik und extraterrestrische Physik, Technische Universität Braunschweig, \\Mendelssohnstr. 3, D-38106 Braunschweig, Germany}
\address[heid]{Institut für Geowissenschaften, Ruprecht-Karls-Universität Heidelberg, \\Im Neuenheimer Feld 234-236, D-69120 Heidelberg, Germany}

\begin{abstract}
We present a non-invasive technique for measuring the thermal conductivity of fragile and sensitive materials. In the context of planet formation research, the investigation of the thermal conductivity of porous dust aggregates provide important knowledge about the influence of heating processes, like internal heating by radioactive decay of short-lived nuclei, e.g. $\rm ^{26}Al$, on the evolution and growth of planetesimals. The determination of the thermal conductivity was performed by a combination of laboratory experiments and numerical simulations. An IR camera measured the temperature distribution of the sample surface heated by a well-characterized laser beam. The thermal conductivity as free parameter in the model calculations, exactly emulating the experiment, was varied until the experimental and numerical temperature distributions showed best agreement. Thus, we determined for three types of porous dust samples, consisting of spherical, $\mathrm{1.5 \, \mu m}$-sized $\rm SiO_2$ particles, with volume filling factors in the range of 15\% to 54\%, the thermal conductivity to be 0.002 to 0.02~\WmK, respectively. From our results, we can conclude that the thermal conductivity mainly depends on the volume filling factor.  Further investigations, which are planned for different materials and varied contact area sizes (produced by sintering), will prove the appropriate dependencies in more detail.
\end{abstract}


\end{frontmatter}

% \linenumbers

\section{Introduction}
In the field of planet-formation research, the investigation of the physical properties of porous dust aggregates -- as analogs for protoplanetary bodies -- is of great importance for understanding the evolution of solid matter in young planetary systems. The evolution of the size (distribution) of dust aggregates depends heavily on the collision properties of these particles (see \citet{ZsomEtal:2010,GuettlerEtal:2010}). For larger dust aggregates, other properties become also relevant. Here, we will discuss the thermal conductivity of macroscopic dust aggregates with various packing densities and try to figure out to what extent heating processes can alter their internal structure. Beside globally elevated temperatures in protoplanetary disks and episodic heating events, internal heating of larger bodies by radioactive decay of short-lived nuclei, e.g. $\rm ^{26}Al$, can lead to significant structural metamorphism of planetesimals \citep{GoepelEtal:1994,TrieloffEtal:2003,KleineEtal:2008,PrialnikPodolak:1999,PrialnikEtal:2008}.

Planetesimals are initially highly porous and fragile bodies. As high porosity bodies have only few contacts between neighboring monomers of the material and/or very small contact areas between them, a lower thermal conductivity is expected than in more compact bodies. If a porous body is heated from inside by radioactive decay, sintering between monomer grains in contact can occur. By the formation of inter-particle necks due to the sintering process, the contact areas between the single particles will be enlarged and thus the material solidifies \citep{Poppe:2003}. The larger contact areas of the sintered material would imply a higher thermal conductivity and as a consequence a more efficient and faster heat transport to the surface of the body. The intimate interplay between heating due to low thermal conductivity and an increase in heat flow caused by sintering is a yet unsolved problem in planet-formation research. The question whether these two counteracting effects lead to a complete or partial melting of the body during growth and how such effects influence the further growth and the structural and physico-chemical evolution of planetesimals are yet to be answered.

Although the thermal conductivity of porous media is an important quantity for heat transfer in many other disciplines, like e.g. geology and engineering (e.g. \citet{vanAntwerpenEtal:2010,MasamuneEtal:1963}), comparable thermal conductivity measurements for highly porous matter do not exist, which vary the bulk porosity and packing structure, use particle sizes in the order of few micrometers, and are performed under vacuum conditions. Most of the literature about heat conduction of particulate materials deals with gaseous or fluid transport through the voids, whereas typical gas pressures in protoplanetary disks or (at a later stage) debris disks are so low that the heat transport through the pore space of dust aggregates by gas molecules can be neglected (see also Sect.~\ref{sec:discussion}). Comparable measurements under vacuum conditions or at low atmospheric pressures found in the literature mostly concentrate on larger particle sizes and higher (and often inhomogeneous) packing densities (e.g. \citet{HuetterEtal:2008,KuehrtEtal:1995,Merrill:1969}).

The presumably most unaltered protoplanetary material available in the Solar System can be found in comet nuclei, and recent infrared measurements of Comet 9P/Tempel 1 have shown that the thermal inertia $I = \sqrt{k \rho C}$, with $k$, $\rho$, and $C$ being the heat conductivity, the density, and the specific heat capacity, respectively, is as small as $I < 50 \, \rm W\,K^{-1}\,m^{-2}\,s^{1/2}$ \citep{GroussinEtal:2007}. Assuming $\rho = 100 \ldots 1,000 \, \rm kg \, m^{-3}$ and $C = 700 \ldots 1,400 \, \rm J\,kg^{-1}\,K^{-1}$, a thermal conductivity of $k < 1.8 \times 10^{-3} \ldots 3.6 \times 10^{-2}$\,\WmK~can be derived. Mind, that the low thermal-inertia values derived by \citet{GroussinEtal:2007} are in conflict with a different analysis of the same data by \citet{DavidssonEtal:2009}, who yield thermal inertias of wide surface ranges of Comet 9P/Tempel 1 of $1000-3000 \, \rm W\,K^{-1}\,m^{-2}\,s^{1/2}$, whereas only small fractions of the comet surface possess low thermal inertias of $40-380 \, \rm W\,K^{-1}\,m^{-2}\,s^{1/2}$. The measurement of the thermal conductivity of porous dusty media can, thus, provide information about the morphology and structure of the surface material of comets. Such results are not only important for the early evolution of small Solar System bodies, but also for thermal processes in evolved systems, e.g. for heating and implicit activity of comets with their porous ice structure and dust mantles during passage through the inner Solar System, or the Yarkovsky effect for asteroids that may have regolith on their surfaces.

The thermal conductivity of porous media is affected by a large range of parameters: the bulk material in general (pure or mixed), the temperature, the porosity, the shape of monomers, the size distribution of the grains (mono- or polydisperse), the number of contact points between the particles, the size distribution of the contact area (potentially changed by sintering effects, see \citet{KossackiEtal:1994,SeiferlinEtal:1995}), and the structure of the bulk matrix. The latter can be disordered or ordered, and in the ordered case can additionally be divided into isotropic (with simple or complex structure) and anisotropic (isotropic in one plane or not isotropic in any plane).

Many models to describe the evolution of planetesimals or parent bodies of meteorites use thermal conductivity values derived by measurements of chondrites \citep{MiyamotoEtal:1981,GhoshMcSween:1998,McSweenEtal:2002,GhoshEtal:2003,GhoshEtal:2006,MerkPrialnik:2003,MerkPrialnik:2006,HeveySanders:2006}. However, this material has already undergone several thermally and mechanically induced structural modifications and, thus, cannot correctly represent the primary highly porous state of these bodies. Even an often used correction or reduction factor for the heat conductivity to account for the porosity or the cohesiveness of the bulk material cannot describe the complex correlation between the thermal conductivity and the inner structure of the material to the full extent (e.g. \citet{CapriaEtal:2002,SeiferlinEtal:1996,PatonEtal:2010}). For many decades, extensive studies have been made to describe the relationship between the thermal conductivity and the porosity of two-phase mixtures and porous materials (see reviews by \citet{ProgelhofEtal:1976}, \citet{ChengVachon:1970}). All the semi-empirical equations mentioned in the literature are based on well-defined material configurations and environmental conditions and thus are restricted to specific application problems. In addition to that, most of these formulae consider the porosity as the most important parameter defining the thermal conductivity without including the influence of pore size, pore shape, material structure, etc..  In this work, the numerical model reproducing our experimental measurements does not include a detailed theoretical description about the thermal conduction of porous material but uses a reduction factor of the thermal conductivity as a free model parameter (see Sect.~\ref{sec:model}), which indirectly but not explicitly includes the manifold structural parameters of the bulk material contributing to the heat conduction.

To understand the intricate dependency between the thermal conductivity and the structure of the protoplanetary body, extensive experimental investigations are needed. The results of these measurements will serve as important material parameters for modeling the internal constitution and thermal evolution of planetesimals and cometary nuclei and generally for other thermal processes such as thermophoresis and photophoresis (e.g. \citet{WurmHaack:2009}).

For investigating the heat conductivity of porous and fragile dust samples, it is very important to choose an appropriate measuring technique. Conventional methods, like the hot wire or the hot plate method (see review by \citet{PresleyChristensen:1997-1,PresleyChristensen:1997-2}), all need direct contact to the material. By establishing the contact between the measuring apparatus and the material of loose dust assemblies, the frail inner structure of the dust sample will always be changed in terms of compression or disruption. Compressive and tensile strengths of our materials are typically in the range of a few 100 to a few 1,000 Pa \citep{BlumEtal:2006}, and even the slightest contact with a heating wire will cause unknown local changes of the aggregate morphology. Thus, we have developed a non-invasive technique to measure the thermal conductivity by heating the dust aggregate with a laser beam and recording the temporal and spatial propagation of the heat wave by an IR camera, which is comparable to the laser flash technique (e.g. \citet{ParkerEtal:1961,FayetteEtal:2000}). The temperature distribution on the surface and in the interior of the sample is modeled using variable thermal conductivity values until the best fit to the experimental results is achieved.

In this first publication about thermal conductivity measurements in the context of planet formation research, we will present the measuring technique and the corresponding model calculations for a set of exemplary dust samples with different porosities and different internal structures, consisting of the same dust type. Further measurements with other protoplanetary relevant materials and diverse sintering stages are under way.

\section{\label{sec:experimental_setup}Experimental Setup}
\begin{figure}[h!]
    \center
    \includegraphics[height=6cm]{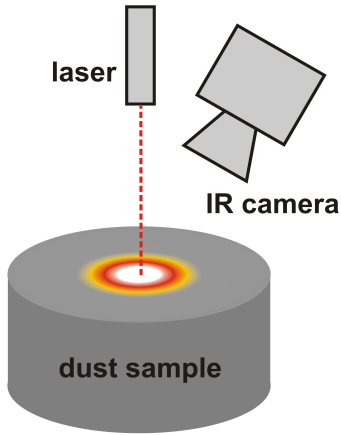}
    \caption{\label{exp_setup}Experimental principle for measuring the thermal conductivity of a cylindrically shaped dust sample. The sample is heated by a laser beam, while an IR camera monitors the temporal and spatial temperature distribution.}
\end{figure}

As we expect very low values for the thermal conductivity of porous dust aggregates in the order of $\sim 10^{-3} \ldots 10^{-2}$\,\WmK, as indicated by the measurements of Comet 9P/Tempel 1 (\citet{GroussinEtal:2007}, see above), high-vacuum conditions are required for the measurements, as the presence of (even rarefied) air would significantly contribute to the thermal conductivity. Hence, all measurements were performed in a vacuum chamber at pressures around $10^{-5}\,\rm mbar$. From outside the vacuum chamber, an infrared laser beam (wavelength~=~813\,nm), an overview camera to align the laser, and an IR camera are pointed through respective windows onto the aggregate inside the chamber. To prevent additional sources for heat conduction, the dust sample is positioned onto a block of insulation material with a thermal conductivity of $\sim 0.022$\,\WmK. The thickness of the dust samples is so large that we do not expect a considerable heat flow to this substrate (see Sect. \ref{sec:results} and Fig. \ref{vertcrosssection}). The cylindrically shaped dust aggregate (see Sect.~\ref{sec:sample prep}) is directly heated by the laser beam mounted vertically to the surface of the dust sample (see Fig.~\ref{exp_setup}).

During the heating and cooling phase, an IR camera monitors the temporal and spatial temperature distribution of the aggregate's surface at an inclination angle of $\sim 30^{\circ}$ to the surface normal. The images of the IR camera have a spatial resolution of $\sim 0.3\,\rm mm$\ and a temporal resolution of $0.02\,\rm s$. The temperature measurements of the IR camera were performed with a default emissivity value of unity. To achieve more accurate temperature values, we calibrated the temperatures of the IR camera by comparison to a heated material with the same IR emissivity value as the dust sample, using a thermocouple for precise direct temperature measurement. Each sample was heated by the laser beam until the temperature of the spot of the laser beam on the sample surface did not considerably change, i.e. the heating by the laser and the conduction of the heat inside the dust sample were reaching an equilibrium state. The cooling phase, after the laser had been switched off, was recorded until the dust material of the sample was cooled down to room temperature.

\section{\label{sec:sample prep}Sample Preparation}
For the first thermal conductivity measurements, we used $\rm SiO_2$ dust, consisting of monodisperse spheres with a diameter of $\mathrm{1.5\, \mu m}$. This material is widely used as protoplanetary dust analog and is well-known to us in terms of physical properties after years of research \citep{BlumEtal:2006,GuettlerEtal:2010}. One additional advantage to choose this material for our initial experiments is the spherical shape of its monomers and their well-defined size. For future numerical simulations, information concerning the internal structure of the sample material is required to model the heat conduction through the contact areas of neighboring individual particles. The geometrically simple shape and structure of this dust material facilitates the development and testing of such a numerical thermal conductivity model.

\begin{figure}[b!]
    \center
    \includegraphics[width=.7\columnwidth]{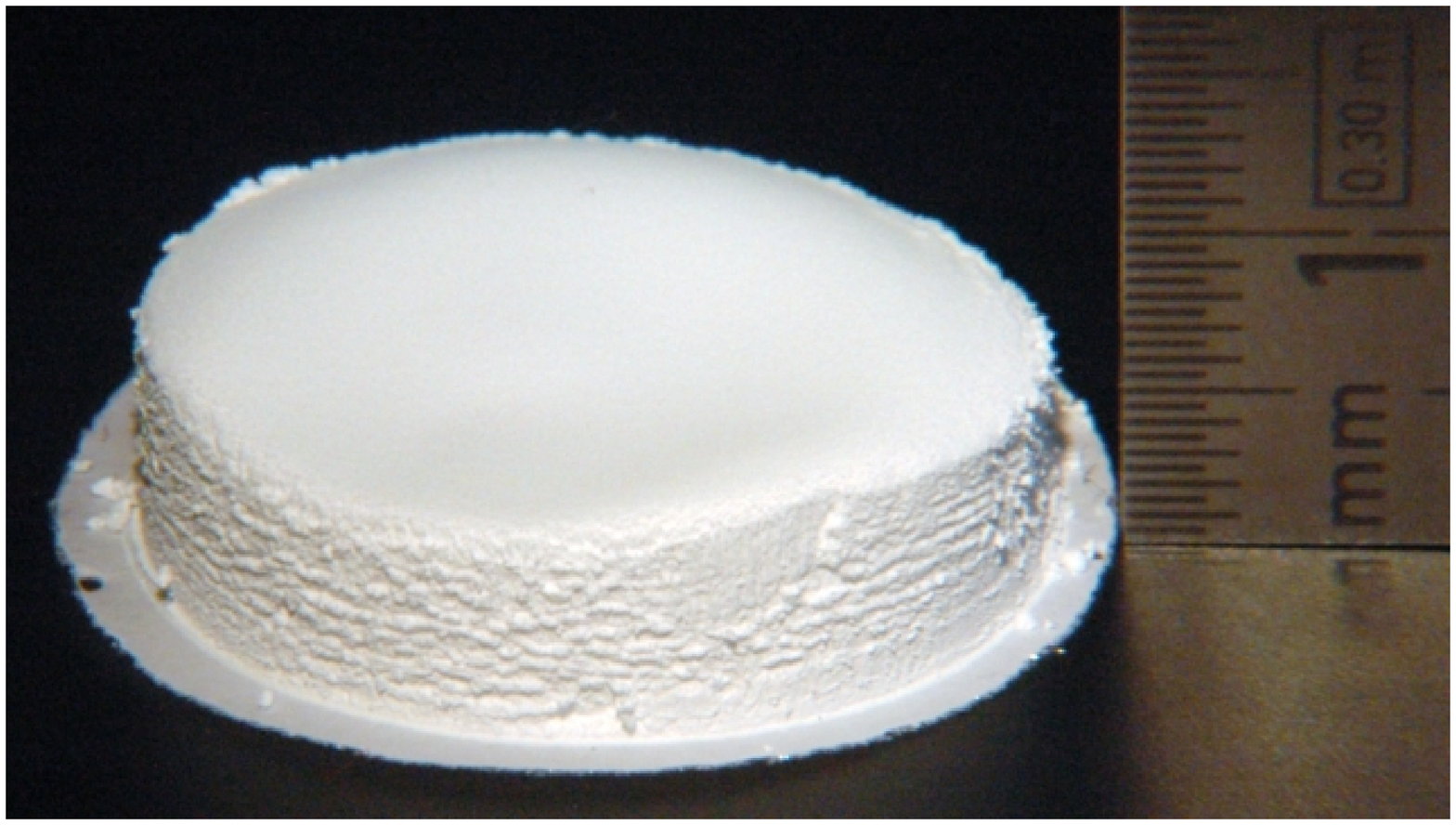}\vspace{5pt}
    \includegraphics[width=.7\columnwidth]{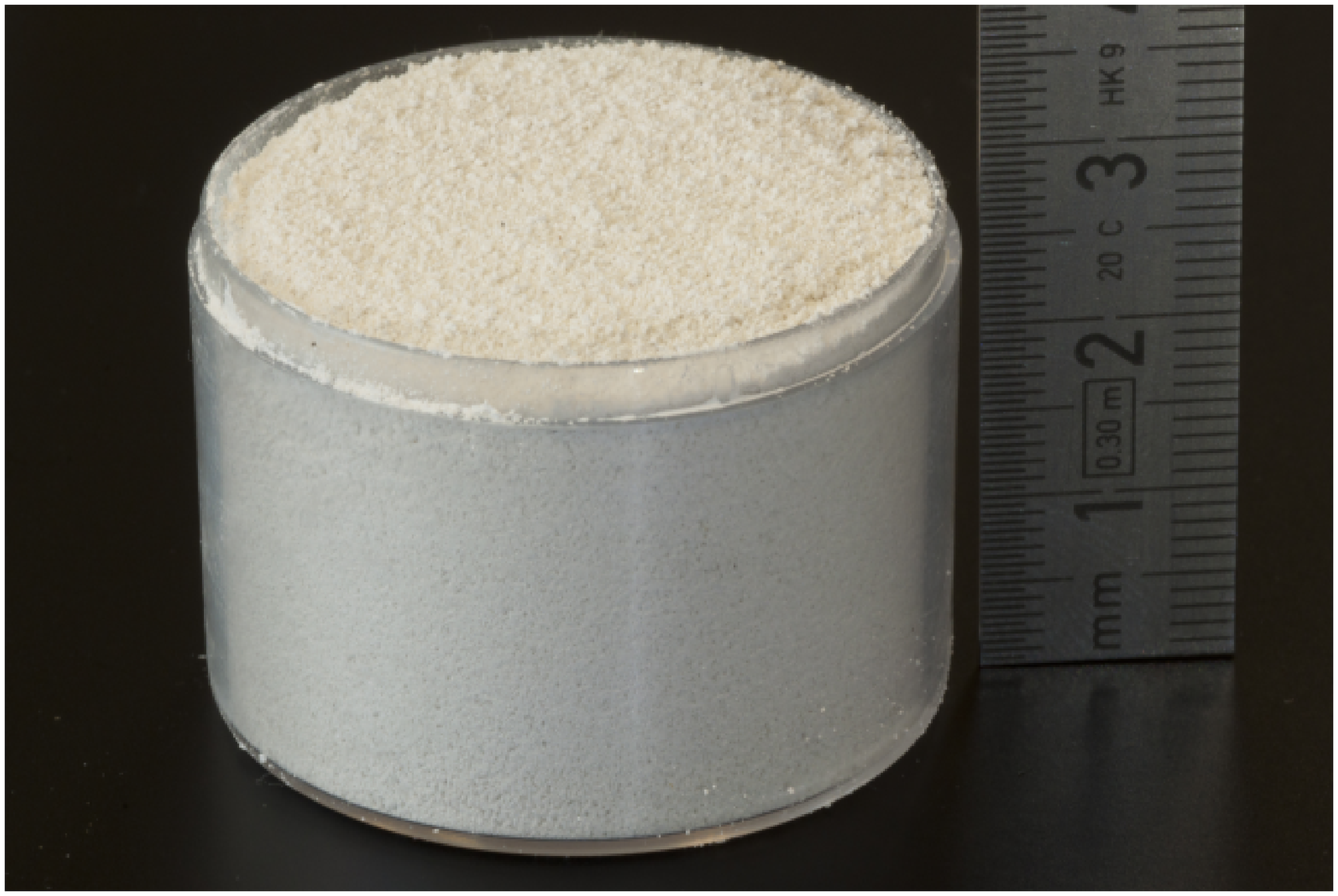}\vspace{5pt}
    \includegraphics[width=.7\columnwidth]{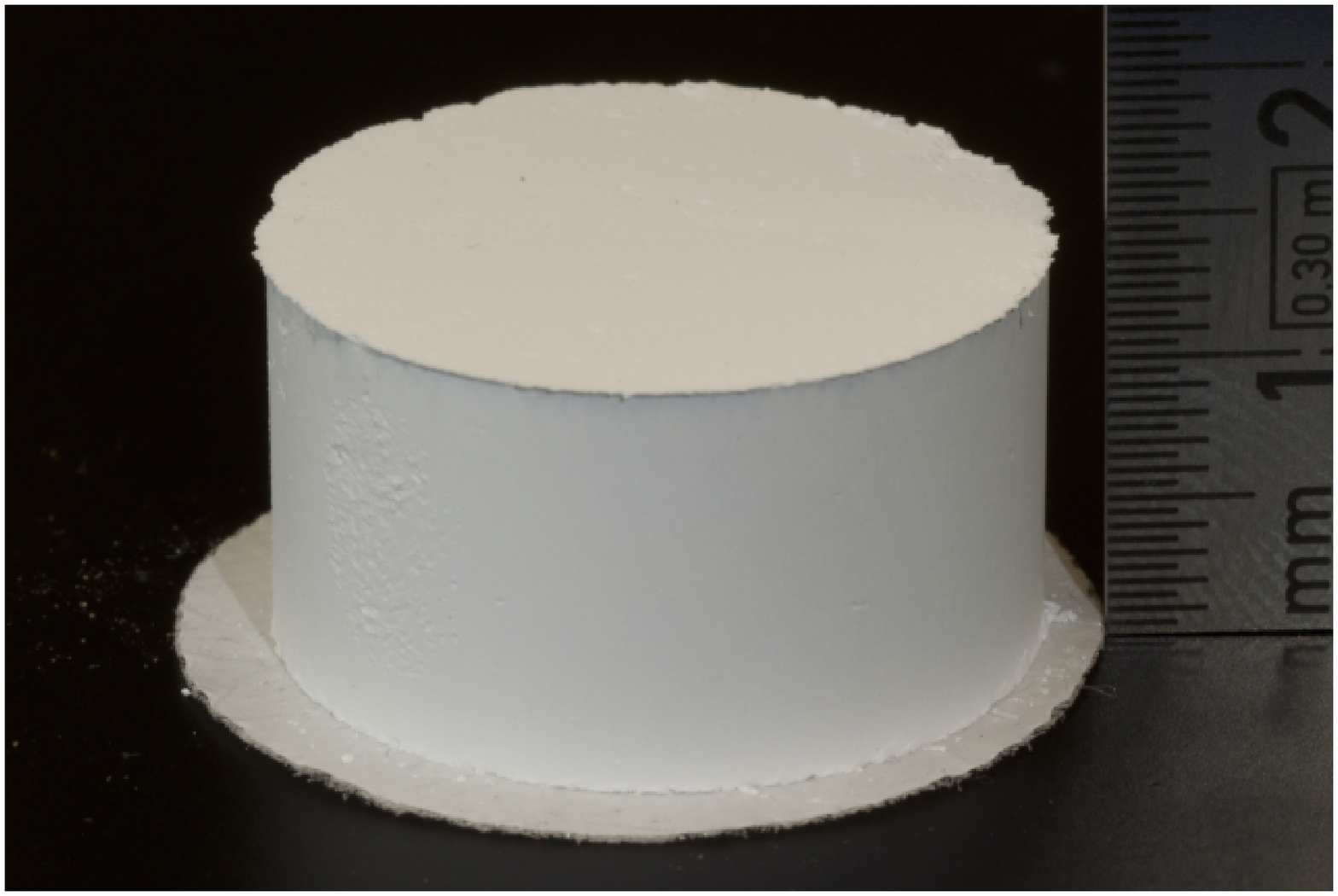}
    \caption{\label{sample_fotos} Examples of dust aggregates manufactured by the three different methods described in the text. Top: dust sample produced by the RBD method; middle: sieved dust sample; bottom: compressed dust sample.}
\end{figure}

To study the influence of the material porosity on the thermal conductivity, we manufactured and analyzed dust samples of different porosities, resulting from three different production techniques. By the random ballistic deposition (RBD) method \citep{BlumSchraepler:2004} we produced highly-porous cylindrically shaped dust samples with a diameter of $2.5\,\rm cm$ and a volume filling factor of $\phi = 0.15$ (see Fig.~\ref{sample_fotos}, top). To achieve moderate volume filling factors in the range of $\phi = 0.15 \ldots 0.30$, we sieved the dust through metal filters with different mesh apertures into a cylindrically shaped plastic container with a diameter of $3.5\,\rm cm$ (see Fig.~\ref{sample_fotos}, middle). The collection of the sieved dust into the container was necessary to achieve well-defined borders of the sample and a flat surface. Previous measurements showed that the propagation of the heat wave for this material is not large enough to reach the borders of the container so that the container material has no influence on the thermal conductivity of the dust sample. More compact samples with a volume filling factor of $\phi \approx 0.5$ were produced by compressing the dust by a plunger inside a hollow cylinder with an inner diameter of $2.5\,\rm cm$ (see Fig.~\ref{sample_fotos}, bottom).

\section{\label{sec:model}Modeling}
The problem to derive the thermal conductivity of a body from measured distributions of surface temperatures can be solved by numerical modeling. The heated sample is described as a straight cylinder with height and radius measured in the experiment, i.e. the solution to the heat conduction equation is found in a two-dimensional axisymmetric geometry. The volume filling factor is defined experimentally, whereas the heat capacity and the density of the solid material are taken from the literature to be 840\,$\rm J\,kg^{-1} K^{-1}$ \citep{Tipler:1999} and 2,000\,$\rm kg\,m^{-3}$ \citep{BlumSchraepler:2004}, respectively. It is assumed that the sample does not contain volatile components and that the ambient pressure is extremely low (because the actual experiment was conducted in a vacuum chamber, see Sect.~\ref{sec:experimental_setup}).

A two-dimensional transient nonlinear heat conduction equation for the porous sample with volume filling factor $\phi$ is solved. Note that we use a continuous description for the porous medium. This means that the classical heat transfer equation is used with the effective parameters (density $\rho ({\rm {\bf r}}) = \rho_0 \, \phi$ and heat capacity $c({\rm {\bf r}}))$, defined for a particular porous medium
\begin{equation}\label{eq:heat_transfer}
\frac{\partial T(t,{\rm {\bf r}})}{\partial t} c({\rm {\bf r}}) \rho_0 ({\rm {\bf r}}) \phi (t,{\rm {\bf r}}) = \nabla (k(T,{\rm {\bf r}}) \nabla T(t,{\rm {\bf r}})) + E(t,{\rm {\bf r}})\;.
\end{equation}
The first term on the right-hand side of Eq.~\ref{eq:heat_transfer} describes the net change of heat in the unit volume due to heat conduction. The effective conductivity $k(T, {\bf r})$ is given by $k(T, {\bf r})=h\, k_{d}+k_{I\!R}$, where the approximate bulk conductivity of a solid mixture of dust, $k_{d}$, has been corrected for the aggregate structure via a reduction factor $h$. This factor is strongly dependent on the morphological structure of the medium, e.g., the monomer size and shape, the porosity, the coordination number, and the sintering state of the dust material. As these parameters are hard to express quantitatively, we treat $h$ as a free parameter in the present work. It is assumed that the conductivity of solid dust $k_{d}$ is constant and equal to 1.4\,\WmK. We also assume that the dust particles are completely opaque in the thermal radiation range. The radiative conductivity $k_{I\!R}$ is taken into account when the total thermal conductivity of the medium is calculated, and is defined as
\begin{equation}
k_{I\!R}(T) = 4 \sigma \varepsilon T^3 l \;,
\end{equation}
with the emissivity $\varepsilon$ equals (1-albedo) according to Kirchhoff's law, $\sigma$ the Stefan-Boltzmann constant, and $l$ the inter-particle distance.

Since all the samples used in our experiments are characterized by low packing densities (the maximum value of the filling factor is equal to 0.5 for the compressed sample), we include bulk absorption of energy in the numerical model. This means that the incident radiation is absorbed in the volume, rather than on the sample surface. We assume that the weakening of the intensity of the direct laser radiation obeys Beer's law, i.e. decreases exponentially with depth. The second term on the right-hand side of Eq.~\ref{eq:heat_transfer} is a source term for the laser energy absorbed by the unit volume
\begin{equation}
E(t,{\rm {\bf r}})=E_0 \exp (-{\rm {\bf r}}/\lambda ) \label{eq:laser_rad} \;,
\end{equation}
here $E(t, {\bf r})$ is the net downward directed energy flux, $E_{0}$ and $\lambda$ are a normalization factor and an attenuation length, respectively.

We do not consider at this stage the effects associated with multiple scattering of direct radiation in the porous layer. The fact is that, although our samples possess significant porosity, this porosity is not sufficient to apply the traditional computational methods of modeling the transport of electromagnetic radiation, designed for sparse environments. Application of the classical equation of radiative transfer can result in noticeable errors already at volume filling factors of $\phi = 0.30$ \citep{Tishkovets:2006}.

The upper boundary condition for the temperature is obtained from the energy balance equation for the surface. Energy is given to the surface by laser radiation and heat conduction (in case of positive temperature gradient). Energy is removed from the surface due to thermal radiation and heat conductivity (in case of negative temperature gradient). Mathematically, the condition is formulated as
\begin{equation}
k \nabla T(t,{\rm {\bf r}}) = \varepsilon \sigma T^4 \;,
\end{equation}
where we have a balance between thermal re-radiation and heat conductivity. We note that the radiation term is absent in the case of porous media. In this way, the heat transfer equation deals with radiation as a source term \citep{DavidssonSkorov:2002}. The thermally insulating boundary condition was applied to all the other boundaries of the model. This simple condition has been chosen after tested simulations clearly demonstrated that the heat wave does not reach the model boundaries during the experiment.

An important part of our numerical simulation is the model of the laser beam, which serves as the sole source of heating energy. In the computer model, the laser beam is characterized by its total power, the spatial profile of the intensity, and the duration of the illumination. All these characteristics were determined experimentally (see Sect. \ref{sec:meas_model_param}).

We now describe the numerical model. Our computer experiments (as well as our laboratory experiments) can be divided into two phases: an active phase, when the sample is irradiated by the laser beam, and a passive stage, after the laser is switched off and the heat (wave) is spreading in the porous material. As an initial condition for the temperature inside the samples, an isothermal temperature distribution is used. This initial temperature is given in accordance with the specific experiment. The boundary condition at the top of the cylindrical sample is the energy balance and, under the above assumptions, this condition contains only a term that defines the thermal radiation of the sample and a thermal flux. At all other boundaries of the sample, the insulation condition is applied.

A non-uniform adaptive grid is used in the numerical simulations. The minimum spatial step is $10\,\mu\rm m$. Usually the numerical grid contains about 30,000 elements. We use UMFPACK, which can solve linear sparse systems via LU factorization \citep{Davis:2004}. This direct solver is applied to the whole matrix of grid points and is efficient for our 2D problem, although it may be too memory-intensive for 3D simulations.

From the point of view of computer modeling, the specific feature of the considered problem is the presence of three free model parameters, i.e. the heat conductivity of the sample, the albedo of the sample at the wavelength of the laser beam, and the penetration depth of the laser beam into the sample. This fact makes the task of determining the effective thermal conductivity more complex than the standard solution of the non-linear transient heat transfer problem in porous media. In fact, our problem can be formulated as the problem of finding the optimal set of model parameters that yields the best agreement between the numerical simulations and the laboratory experiment.

To solve this complex problem, we essentially relied on the setup of the laboratory experiment, namely its division into an active (heating) and passive (cooling) stage. In the first phase, an inhomogeneously heated region is formed in the sample. The particular temperature distribution in this zone is defined by the absorption of the laser energy as well as by the heat transfer due to contact and radiative conduction. During the cooling stage, the radiation source is turned off, and the evolution of the heated zone is determined only by the effective thermal conductivity of the medium. This fact makes the task of the derivation of the thermal conductivity easier. Thus, initially the problem of the quantitative evaluation of the model parameters $E_0$ and $\lambda$ is solved. The experimental estimates for the radiation absorption and for the size of the area in which almost all the radiation is absorbed (see Sect. \ref{sec:meas_model_param}), are used as initial approximations in these simulations. The temperature evolution during the active phase is characterized by the first and second time derivatives of the temperature and by the maximum surface temperature achieved at the end of the irradiation phase. These characteristics, derived from the laboratory and the computer experiments, are compared for a few selected values of the distance from the axis of symmetry of the sample. Then, using the calculated temperature distribution obtained at the end of the heating phase as an initial temperature distribution for the further simulations, we treat the thermal conductivity as the only free parameter and make a fine adjustment, looking for the best compliance between the computer model and the laboratory experiments.

\subsection{\label{sec:meas_model_param}Measurements for Model Parameters}
To reproduce the laboratory experiments by the numerical simulations in the most realistic way, several additional measurements were necessary. The volume filling factors of the different dust samples were determined using the relation $\phi = m/(V \rho_{0})$ between the mass $m$ and the volume of the dust sample $V$. Here, $\rho_{0} = 2,000\,\rm kg\,m^{-3}$ \citep{BlumSchraepler:2004} is the density of the solid material. In the case of the sieved dust samples, the volume filling factor of the bulk sample $\phi_{bulk} = \phi_{agg}\, \phi_{aa}$ comprises the volume filling factor of the individual sifted (sub-mm-sized) aggregates, $\phi_{agg}$, and the volume filling factor of the ``aggregate of aggregates'', $\phi_{aa}$, which forms when the sifted aggregates fall into the repository. Since it is very difficult to measure the porosity of single very small particles with a diameter in the order of the mesh apertures  $a_{mesh}~=~0.15 \ldots 0.50\,\rm mm$, we estimated these volume filling factors as shown in Table~\ref{porosities_sieved}.
\begin{table}[h!]
\center
\caption{\label{porosities_sieved}Volume filling factors and packing densities of the sieved dust samples.}
\begin{tabular}{lccc}
\toprule
\boldmath{$a_{mesh}$} [mm] & \boldmath{$\phi_{bulk}$} & \boldmath{$\phi_{agg}$} & \boldmath{$\phi_{aa}$}\\
& \footnotesize{(measured)} & \footnotesize{(estimated)} & \footnotesize{(estimated)}\\
\midrule
 0.50 & 0.29 & 0.45 & 0.64 \\
 0.25 & 0.24 & 0.45 & 0.53 \\
 0.15 & 0.16 & 0.45 & 0.35 \\
 \bottomrule
\end{tabular}
\end{table}

As the mesh widths are not too different in size, we assume an identical volume filling factor of $\phi_{agg} = 0.45$ for the sieved agglomerates. This value results if we suppose for the largest sieved agglomerates a packing density of $\phi_{aa} = 0.64$ , which corresponds to a random close packing of equal-sized spheres. In fact, the sieved dust aggregates are not perfectly spherically shaped and do not exhibit a monodisperse size distribution, what would yield a somewhat higher packing density than $\phi_{aa} = 0.64$ \citep{SchaertlSillescu:1994}. In contrast to this, the sieving process of dust aggregates leads to a somewhat less dense packing than the random close packing of solid spheres, due to non-negligible cohesion between neighboring aggregates. As we think that these contrary aspects will probably compensate each other, we decided for a packing density of $\phi_{aa} = 0.64$ for the largest sieved agglomerates as an upper limit. In any case, a filling factor of $\phi_{agg} = 0.45$ is in agreement with measurements of dust aggregates sifted with larger mesh sizes (R. Weidling, pers. comm.).

The heating process by the laser beam is characterized within the simulation model by the total laser power, the spatial energy distribution within the laser beam, the albedo of the sample, the penetration depth of the laser beam, and the duration of illumination, which we all empirically derived or estimated. Regarding the laser beam, we measured its profile and effective power for different input voltages. Due to an adapted collimation optics, the laser profile deviates from a gaussian profile at the center and by wider flanks; the full width at half maximum power of the laser profiles was determined to be $\sim~1.2\,\rm mm$. To determine how deep the laser beam penetrates into each dust sample type during the heating, we investigated the transmission of the laser light through dust layers of different thicknesses. We found that for the RBD samples, a sample thickness of 1~mm attenuates the laser intensity to $\sim 10^{-2}$ of its original value; for the sifted samples, the attenuation after penetrating a sample thickness of 1~mm was $\sim 10^{-4}$ and for the compressed samples $\sim 10^{-6}$.
These results verify that for all sample types after a depth of a few millimeters, absolutely no light is transmitted anymore.

To estimate the amount of laser light absorbed by the samples, we made use of specific albedo measurements performed at the PTB in Braunschweig. These albedo measurements were restricted to the compressed samples due to the vertical orientation of the sample holder and yielded a value of 0.983 for the albedo. For such high albedo values, however, the measurement uncertainties are quite high. For the more porous samples, the albedo estimation is technically even more challenging. In addition, the increase of porosity can lead to significant variations in albedo associated primarily with the local inhomogeneity of the surface. Thus, we use the measured albedo values for an indication of very high albedos and leave the albedo as a free parameter in the model simulations. It is again important to note that our computer modeling is naturally divided into two stages: the stage in which the sample is heated by the laser beam (active phase), and the stage when the sample cools down (passive stage). In the second stage, the free model parameters albedo and absorption profile of the radiation inside the sample play no direct role. However, the corresponding time evolution of the heated sample depends on the volumetric distribution of absorbed laser energy and, thus, the temperature at the end of the active phase. Therefore, albedo and the attenuation profile of the laser radiation have an indirect influence on the subsequent thermal evolution of the sample. As will been seen in Sect.~\ref{sec:results}, the values of the free model parameters used in our simulations are in the expected range of the available experimental estimates (both for albedo and for the penetration depth of the laser beam). Good quantitative agreement between the results of our computer simulation in the active stage of the experiment with laboratory data can be treated as an additional proof that the required temperature distribution inside the sample has been calculated accurately.

\section{\label{sec:results}Results}
With the procedure described in the previous Sections, we derived the thermal conductivity of the three different dust sample types described in Sect.~\ref{sec:sample prep}, all consisting of the same $\mathrm{1.5 \, \mu m}$-sized $\rm SiO_2$ dust. For all dust samples we recorded the temperature distribution of the sample surfaces by the IR camera for ten different laser intensities. Exemplary, we present the results of two compressed dust samples, three sieved dust samples, each produced by a mesh with a different aperture size, and two dust samples made by the RBD method. In Table~\ref{sample_param}, each numbered sample is listed along with the respective volume filling factor $\phi$ and the used laser power $P_{laser}$.

\begin{table}[!htb]
\center
\caption{\label{sample_param}Parameters of the three different dust sample types: compressed (C), sieved (S), and produced by the RBD method (RBD). Here, $a_{mesh}$ denotes the mesh aperture, $\phi$ the volume filling factor of the bulk sample, and $P_{laser}$ the laser power.}
\begin{tabular}{p{.11\columnwidth} p{.37\columnwidth} cc}
\toprule
\textbf{no.} & \textbf{sample} & \boldmath{$\phi$} & \boldmath{$P_{laser}$} [W]\\
\midrule
 C1 & compressed & 0.54 & 1.9 \\
 C2a & compressed & 0.50 & 1.0 \\
 C2b & compressed & 0.50 & 1.9 \\
 \midrule
 S1 & sieved,\newline $a_{mesh}\,=\,0.50\,\rm mm$ & 0.29 & 1.0 \\
 S2 & sieved,\newline $a_{mesh}\,=\,0.25\,\rm mm$ & 0.24 & 1.0 \\
 S3 & sieved,\newline $a_{mesh}\,=\,0.15\,\rm mm$ & 0.16 & 1.0 \\
 \midrule
 RBD1 & RBD & 0.15 & 1.9 \\
 RBD2 & RBD & 0.15 & 1.9 \\
 \bottomrule
\end{tabular}
\end{table}

\begin{figure}[!htbp]
    \center
    \includegraphics[angle=90,width=1.\columnwidth]{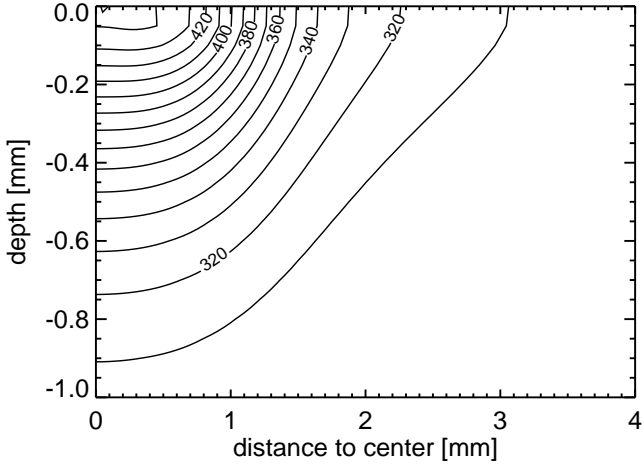}
    \caption{\label{vertcrosssection} Vertical cross section through the temperature distribution inside the modeled sample S1 after a heating duration of 9.5~s. The isotherms are plotted with a division of 10~K.}
\end{figure}

As an example of the result of the numerical model, Fig.~\ref{vertcrosssection} shows the temperature distribution in a  vertical cross section through sample S1 at the end of the heating phase. It is readily visible that the heat wave penetrates to depths of about 1\,mm, which is much less than the thickness of the samples. The radius of the heated zone exceeds its depth by about a factor of three, due to the diameter of the laser beam.

\begin{figure}[!t]
    \center
    \includegraphics[width=1.\columnwidth]{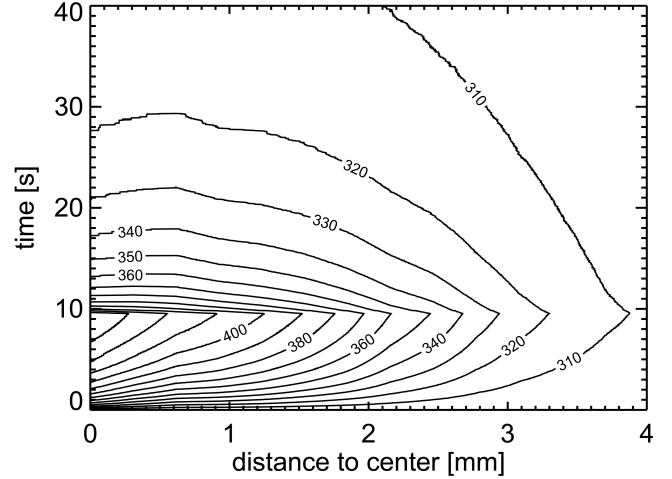}\vspace{5pt}
    \includegraphics[width=1.\columnwidth]{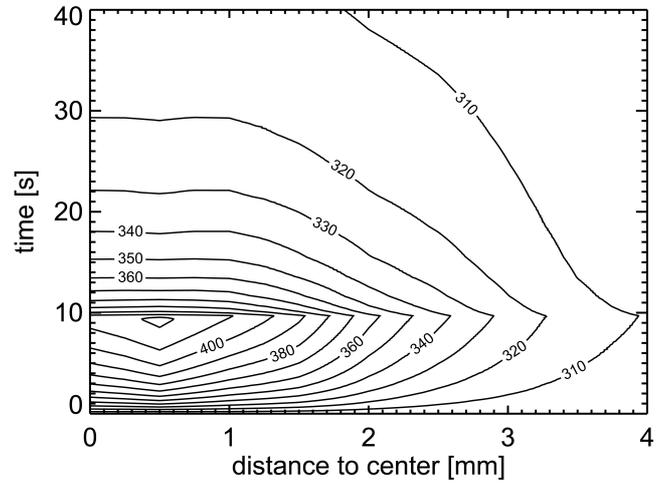}
    \caption{\label{contour_isotherms} Comparison between experimental (top) and numerical (bottom) results for the temporal and spatial temperature evolution of sample S1. The contours display the heating (0 $\ldots$ 9.5~s) and the cooling phase (9.5 $\ldots$ 40~s) of the sample. The temperature intervals of the isotherms correspond to 10~K.}
\end{figure}

Fig.~\ref{contour_isotherms} depicts the measured temporal and spatial surface temperature evolution as a contour plot for sample S1 (top) in comparison to the respective simulated result (bottom). Both diagrams were made by averaging the two-dimensional surface temperature data in 1~pixel-wide annuli around the center of the laser spot. The heating phase during the first $9.5\,\rm s$ can clearly be distinguished from the cooling phase thereafter. Apart from minor deviations, both figures demonstrate the good compliance of measurement and simulation. The deviations in the active phase for distances $<~0.5\,\rm mm$ of the numerical contour plot correspond to the non-gaussian shape of the laser profile.

\begin{figure}[!htbp]
    \center
    \includegraphics[angle=90,width=1.\columnwidth]{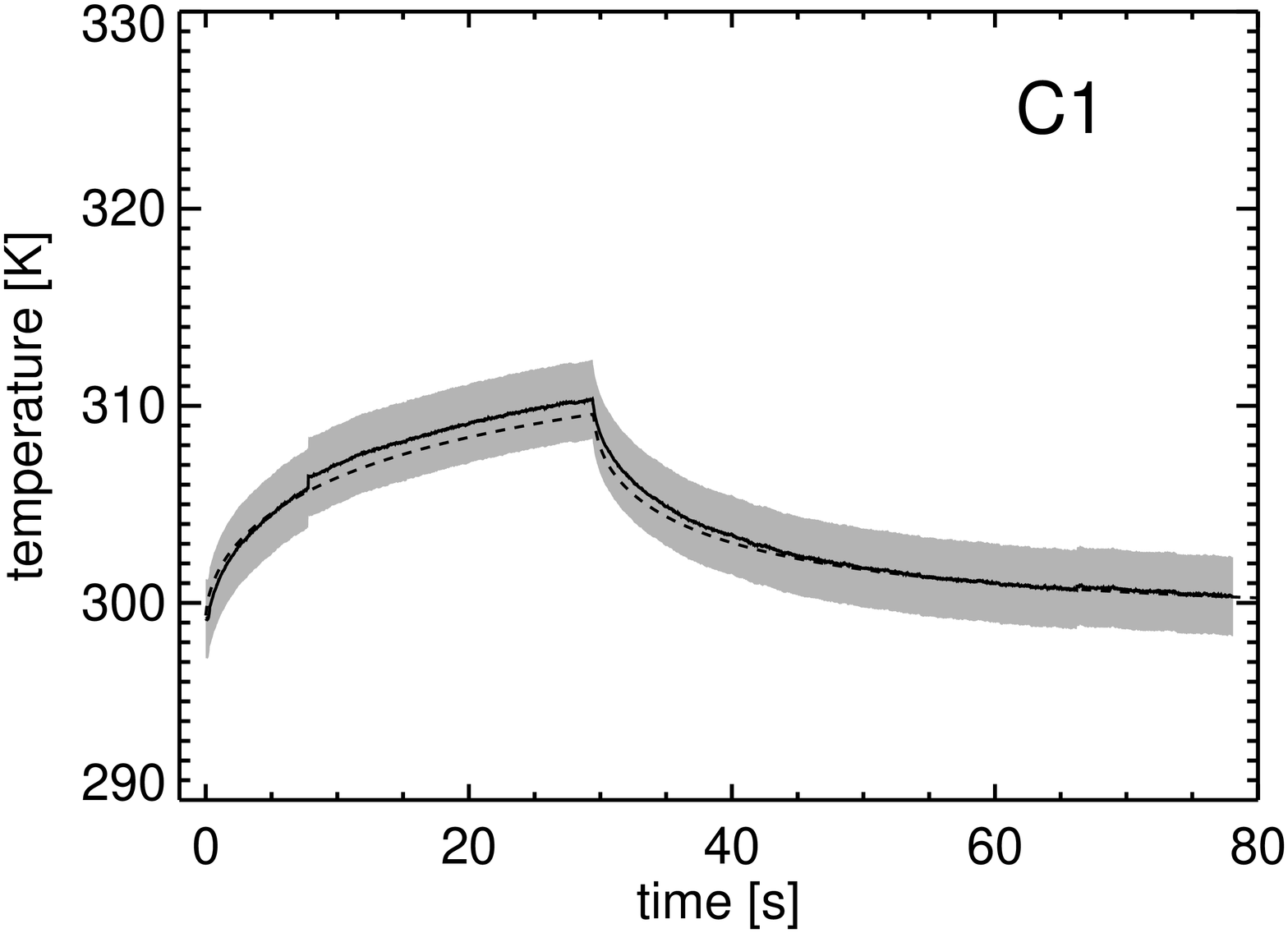}
    \includegraphics[angle=90,width=1.\columnwidth]{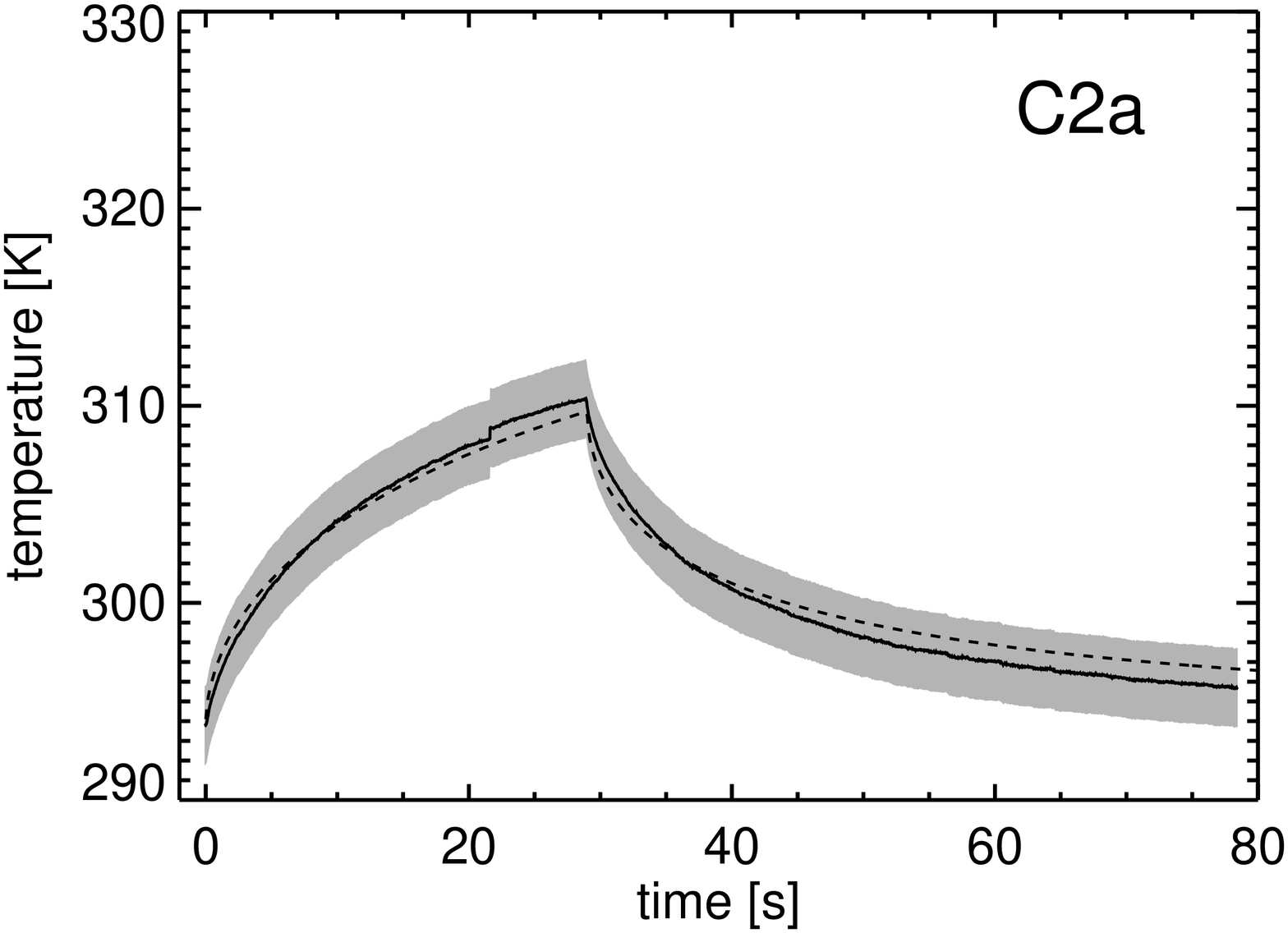}
    \includegraphics[angle=90,width=1.\columnwidth]{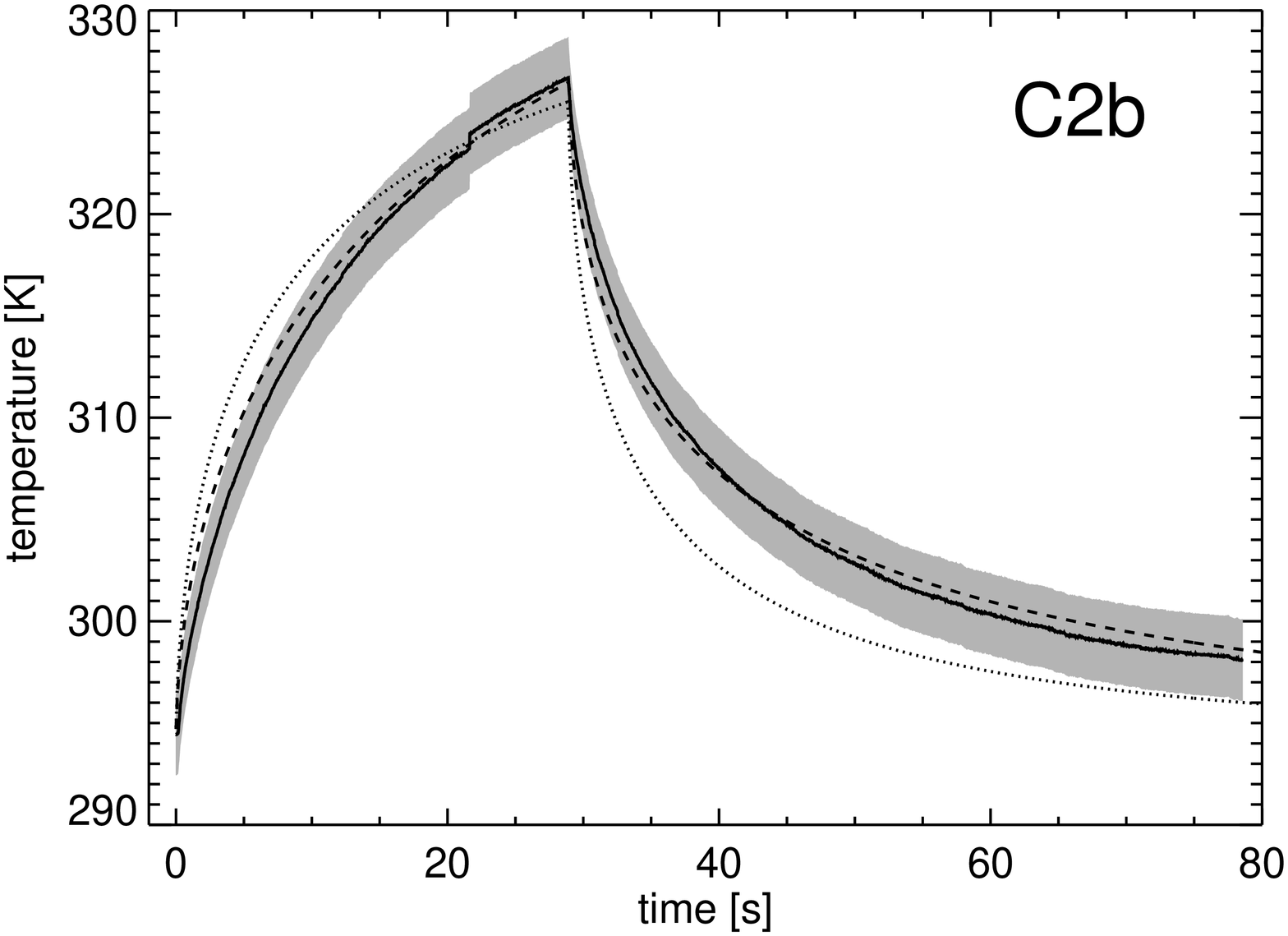}
    \caption{\label{T_t_compressed} Temporal temperature evolution of the compressed dust samples C1, C2a, and C2b, showing the heating and cooling phase for a given distance of $1\,\rm mm$ to the center of the laser beam. The solid lines correspond to the results of the experimental measurements and the dashed lines represent the model calculations. The gray shaded area denotes the errors of the experimental temperature data. The dotted line for sample C2b shows the best-fitting model for the measured albedo of 0.983 and the maximum allowed penetration depth of 0.1\,mm.}
\end{figure}

\begin{figure}[!htbp]
    \center
    \includegraphics[angle=90,width=1.\columnwidth]{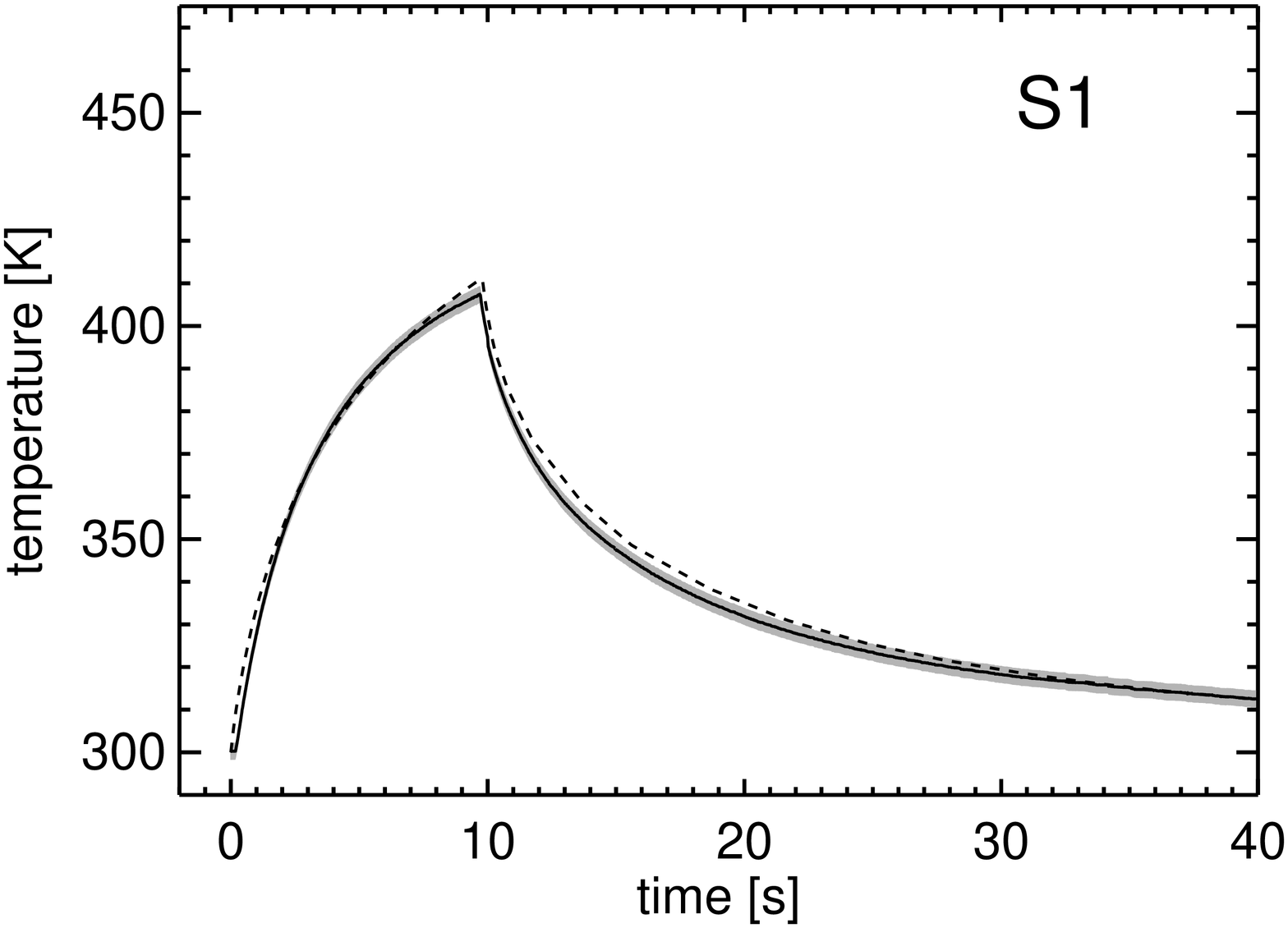}
    \includegraphics[angle=90,width=1.\columnwidth]{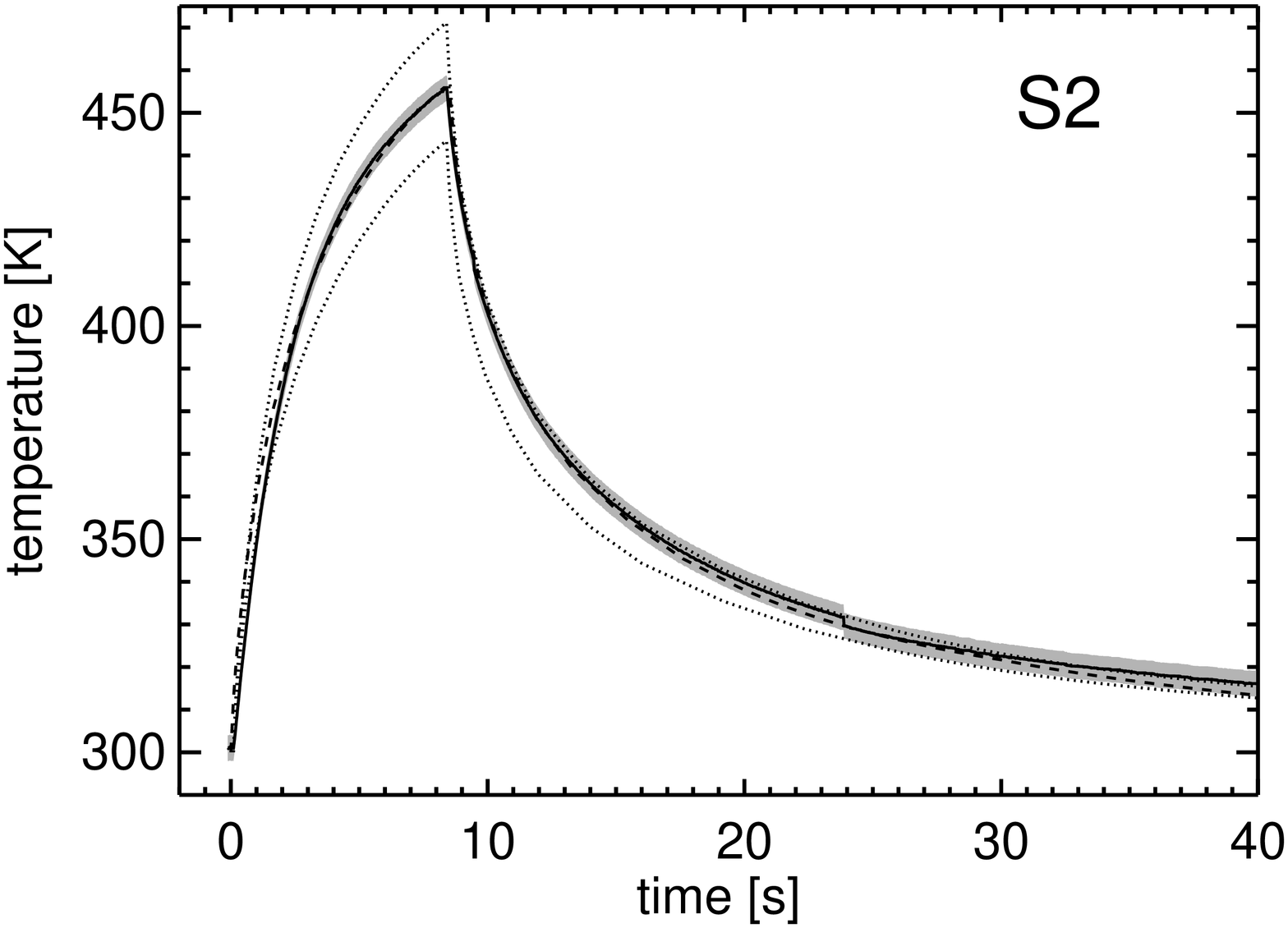}
    \includegraphics[angle=90,width=1.\columnwidth]{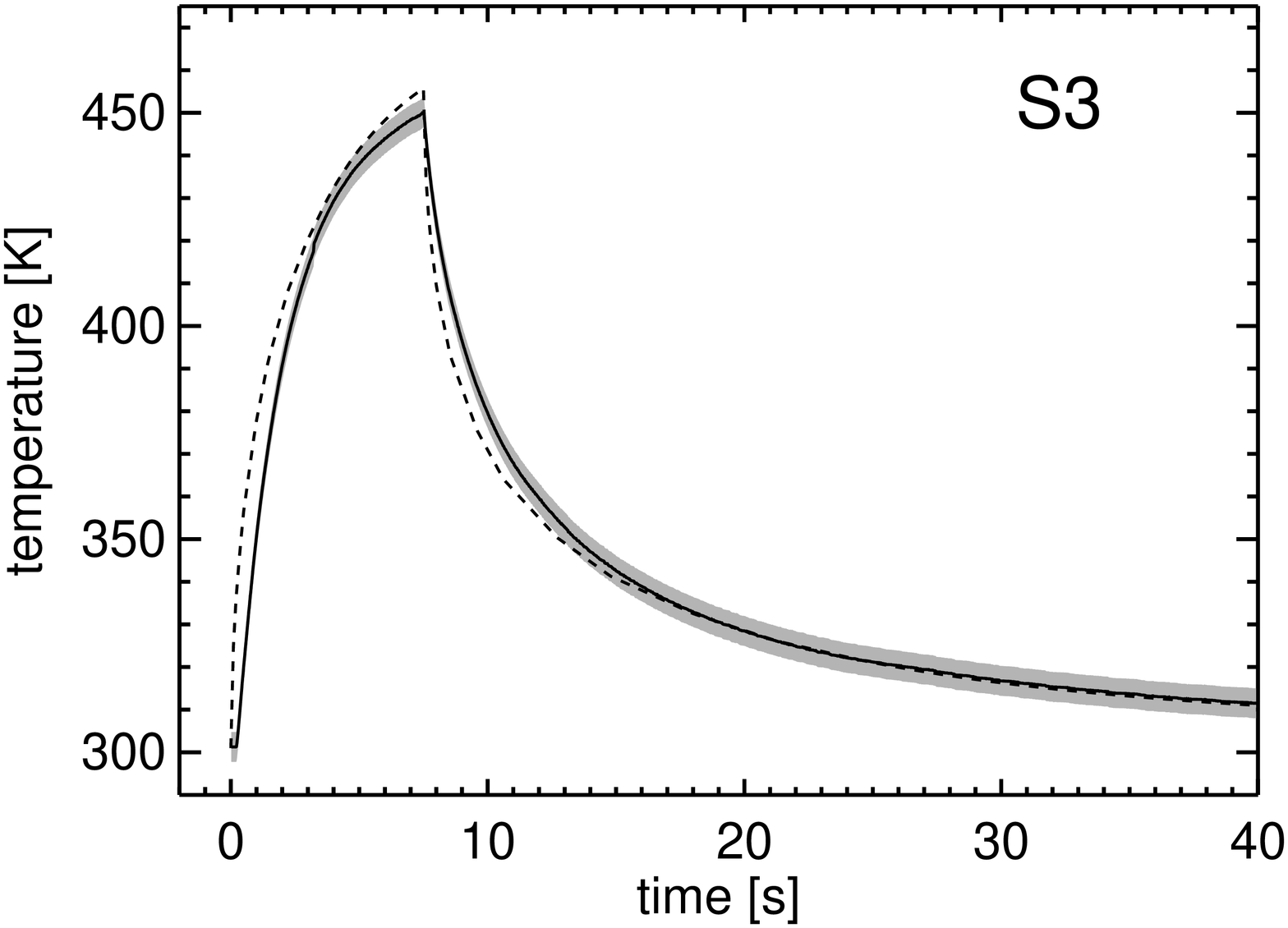}
    \caption{\label{T_t_sieved} Temporal temperature evolution of the sieved dust samples S1, S2, and S3, showing the heating and cooling phase for a given distance of $1\,\rm mm$ to the center of the laser beam. The solid lines correspond to the results of the experimental measurements and the dashed lines represent the model calculations. The gray shaded area denotes the errors of the experimental temperature data. Additionally, the dotted lines in the diagram of sample S2 illustrate the numerical results achieved by a varied thermal conductivity value of $\pm$\,43\% of the best fitting curve.}
\end{figure}

\begin{figure}[!t]
    \center
    \includegraphics[angle=90,width=1.\columnwidth]{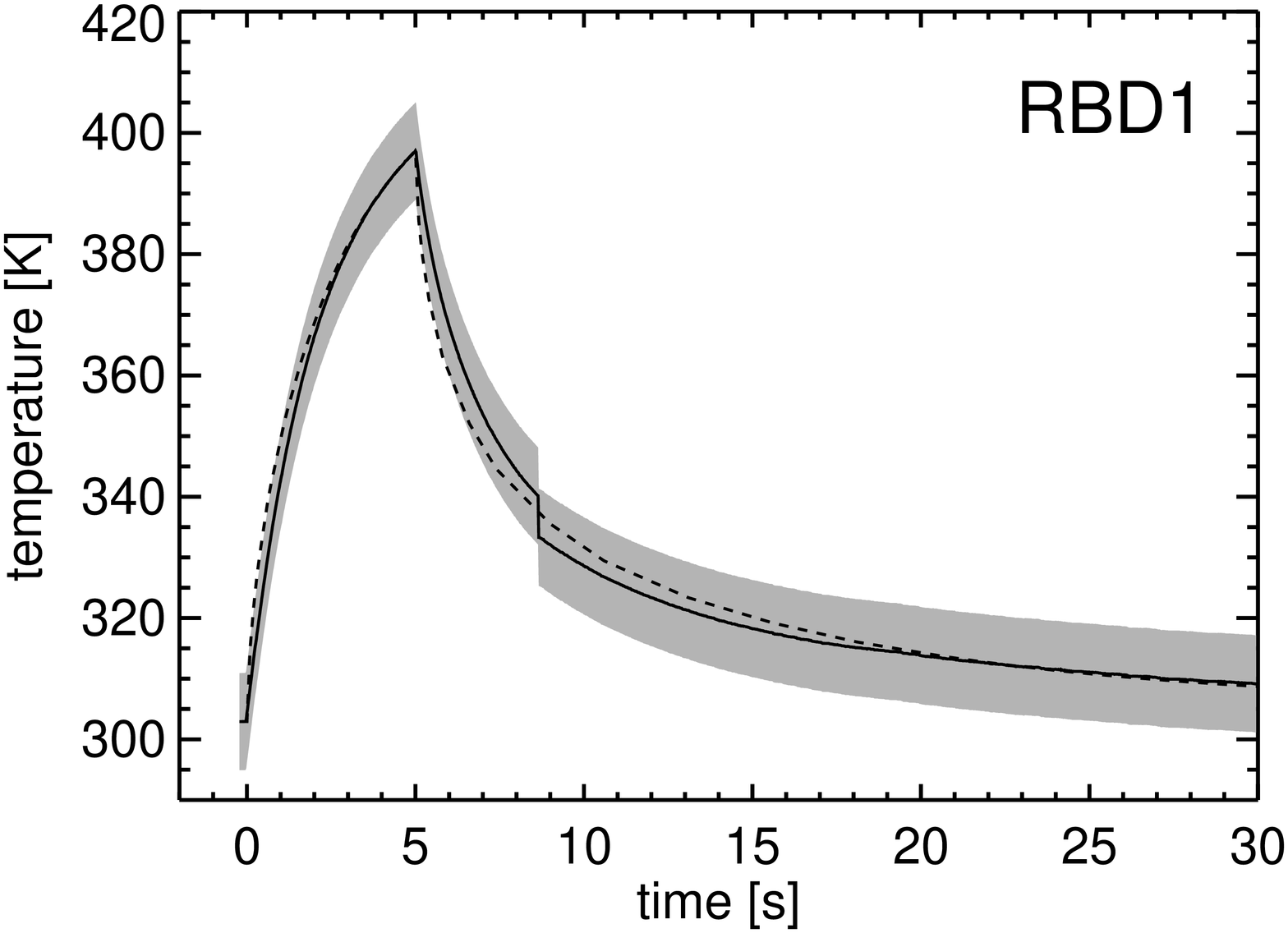}
    \includegraphics[angle=90,width=1.\columnwidth]{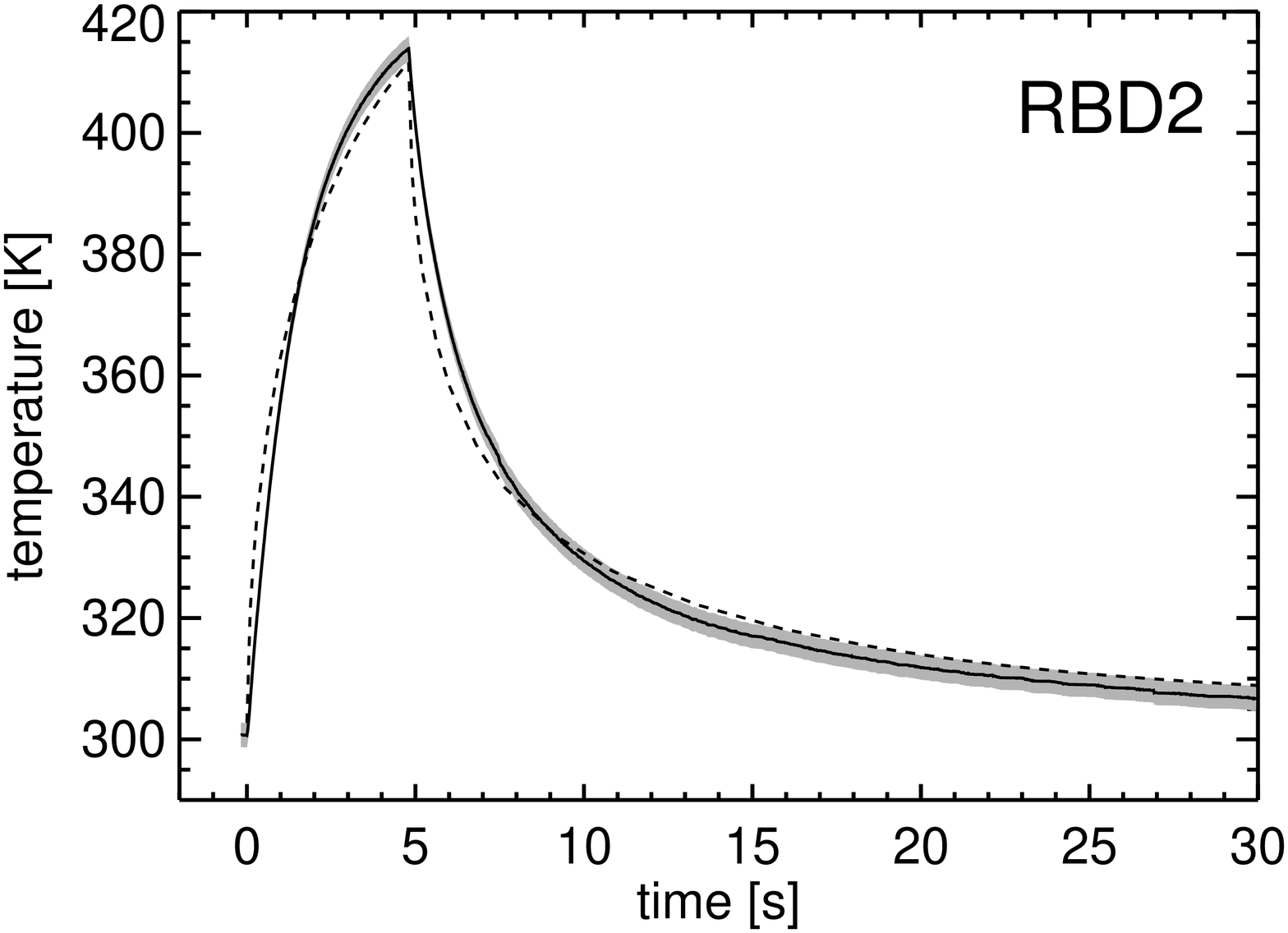}
    \caption{\label{T_t_RBD} Temporal temperature evolution of the dust samples RBD1 and RBD2, produced by the RBD method, showing the heating and cooling phase for a given distance of $1\,\rm mm$ to the center of the laser beam. The solid lines correspond to the results of the experimental measurements and the dashed lines represent the model calculations. The gray shaded area denotes the errors of the experimental temperature data.}
\end{figure}

In Figs.~\ref{T_t_compressed}, \ref{T_t_sieved}, and \ref{T_t_RBD}, the temperature evolution at a distance of $\rm d = 1 \,mm$ from the axis of the laser beam is shown for all eight samples listed in Table \ref{sample_param} to visualize and analyze the congruence of the experimental (marked as solid lines) and numerical (marked as dashed lines) data in detail. These temperature curves correspond to a vertical cut through Fig.~\ref{contour_isotherms} at a distance to the center of 1~mm. The dashed lines in the figures show the best fitting simulated temperature curves, achieved by systematically varying the value of the thermal conductivity for the cooling phase. The other two free parameters, i.e. the albedo of the samples and the extinction depth of the laser beam, were only slightly varied during the heating phase with respect to the values estimated by experimental measurements (see Sect.~\ref{sec:meas_model_param}). The experimentally derived temperature curves show at few positions temperature jumps of up to several degrees K. This effect arises from the self-calibration of the IR camera to correct a time and temperature dependent internal temperature drift. The errors of the experimental temperature data, displayed as gray shaded areas, are defined by the absolute accuracy of the IR camera of $\pm~2\,\rm K$ and by the size of the temperature calibration steps, depending on the value which is dominating. Lying centered within the error intervals of the measured temperature curves, the simulated temperature curves indicate a very good agreement with the experimental data. For comparison, the dotted line for sample C2b in Fig.~\ref{T_t_compressed} shows the model results for the measured albedo of 0.983 and the maximum allowed penetration depth of 0.1\,mm. The resulting thermal conductivity of $k = 0.078$\,\WmK is obviously much too high to fit both the heating and the cooling curve.

To show the sensitivity of the simulated temperature curve on the thermal conductivity as a free parameter in the model, the middle Fig.~\ref{T_t_sieved} displays two additional (dotted) curves with a thermal conductivity value of 0.0052\,\WmK~lying above and 0.0021\,\WmK~lying below the best compatible curve with $k = 0.0036$\,\WmK.

The agreement between of the experimental and modeled temperature curves in Figs.~\ref{T_t_compressed}, \ref{T_t_sieved}, and \ref{T_t_RBD} provides on the one hand a quantitative proof of the validity of the model, but serves on the other hand for a qualitative analysis of the thermal conductivity of each sample. The shape of each individual temperature curve comprises some characteristics that argue independently for a high or low thermal conductivity of the sample. The following quantities suggest a low thermal conductivity in comparison to a sample with a higher thermal conductivity: 1.) The slope of the temperature curve at the start of the heating phase is steep, due to the inability of the sample to carry away the absorbed energy, 2.) the temperature maximum has a higher value for a fixed energy flux, and 3.) the temperature curve has a shallower slope at the start of the cooling phase for a given temperature increase during the heating phase, due to the same argument as in 1). If these characteristics are not fulfilled, a rather high thermal conductivity value is indicated. As the temperature maximum is mainly defined by the absorbed laser power, one should mind to compare the maximum values only between samples heated with the same laser intensity.

\begin{figure}[!htbp]
    \center
    \includegraphics[angle=90,width=1.\columnwidth]{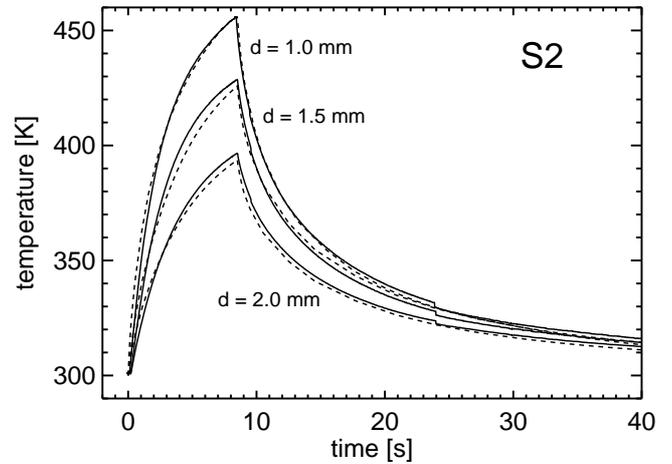}
    \caption{\label{S2_diffdist} Temporal temperature evolution of the sieved dust sample S2 for three distances (1, 1.5 and 2~mm) to the center of the laser beam. The solid lines correspond to the results of the experimental measurements and the dashed lines represent the model calculations. Mind that the thermal conductivity of sample S2 was derived at a distance of 1~mm and still describes the temperature evolution at larger distances quite satisfactorily.}
\end{figure}

The comparison between the experimental and numerical derived data of sample S1 in Fig.~\ref{contour_isotherms} and the temporal temperature evolution of sample S2 (see Fig.~\ref{S2_diffdist}), depicted for three distances to the center of the laser beam (d~=~1, 1.5, 2~mm), represent exemplary the stability of the simulated temperature distribution in temporal as well as in spatial direction. Mind that the thermal conductivity of sample S2 as shown in Fig.~\ref{S2_diffdist} was derived at a distance of 1~mm and still describes the temperature evolution at larger distances quite satisfactorily.

In Table~\ref{sample_results}, the results for the thermal conductivity values of the eight samples are listed, following from the best fitting simulated data. As initially expected, all samples with relatively low volume filling factors of $\phi \approx 0.15$ have a thermal conductivity in the order of $10^{-3}$\,\WmK, whereas the samples with the highest $\phi$ values ($\phi \approx 0.50$) have an order of magnitude higher values for the thermal conductivity.

\begin{table}[!htbp]
\center
\caption{\label{sample_results}Values of the model calculations for the thermal conductivity $k$, the albedo, and the extinction length $\lambda$, which provide the best agreement with the experimentally derived temperature evolution of the eight dust samples in comparison to their volume filling factors $\phi$.}
%\begin{tabular}{p{.15\columnwidth} p{.1\columnwidth} p{.3\columnwidth} p{.15\columnwidth} p{.3\columnwidth} }
\begin{tabular}{lcccr}
\toprule
\textbf{sample} & \boldmath{$\phi$} & \boldmath{$k$} [\WmK] & \textbf{albedo} & \boldmath{$\lambda$} [$\rm \mu$m]\\
\midrule
 C1 & 0.54 & 0.021 & 0.9960 & 50\\
 C2a & 0.50 & 0.021 & 0.9935 & 50\\
 C2b & 0.50 & 0.016 & 0.9945 & 50\\
 \midrule
 S1 & 0.29 & 0.0078 & 0.965 & 50\\
 S2 & 0.24 & 0.0036 & 0.980 & 100\\
 S3 & 0.16 & 0.0016 & 0.972 & 100\\
 \midrule
 RBD1 & 0.15 & 0.0026 & 0.992 & 150\\
 RBD2 & 0.15 & 0.0026 & 0.990 & 100\\
 \bottomrule
\end{tabular}
\end{table}

An estimate of the accuracy with which the thermal conductivity is measured in our work is a non-trivial task. We can confidently discuss the accuracy of the temperature fields measured in the laboratory experiments and the accuracy of the computer simulations. However, both do not completely determine how accurately we estimated the thermal conductivity. The development of a statistical algorithm of accuracy estimation for the proposed technique is beyond the scope of this work. At present, we use the empirical evaluation approach described below.

Let us first quantify the allowed range of the model parameters and their relative errors. We begin with the estimates of the albedo. As previously mentioned, the albedo of the compressed sample was experimentally determined. The measured value is only 2\% smaller than our model value (see Table~\ref{sample_results}). However, it is important to note that it is the albedo (i.e., the intensity of the scattered radiation) that was measured, whereas the absorbed energy, directly proportional to (1-albedo) plays a decisive role in our computer modeling of the heat transfer. It is obvious that for such a high albedo, even minor changes dramatically alter the amount of absorbed energy and, consequently, the temperature evolution of the sample (under the condition that the other model parameters are fixed). The second free parameter of the model is the attenuation factor of the laser radiation (e-folding scale). Accurate measurements of the attenuation of radiation in porous media is a very difficult task. However, we can get some quantitative estimates based on the experimental evaluation of the transparency of our samples. Our estimates show that for the compressed samples, even for a sample thickness of a small fraction of a mm, no more than a few percent of the absorbed energy passes through. This gives us an upper limit of the attenuation factor of $\sim 0.1\,\rm mm$.

Firstly, we tested the possibility to find a better agreement between experiment and computer model using the {\it measured} values of the albedo. As the amount of the absorbed energy is much larger in this case, one has to increase the value of the effective thermal conductivity in the model in order to fit the experimental data. Based on the law of conservation of energy, one could expect that the new value of the coefficient of thermal conductivity should be roughly three times larger than the value presented in the Table~\ref{sample_results}. Our attempts to find a good match between measured and calculated temperature curves show that for all physically reasonable model parameters (attenuation factor $< 0.1\,\rm mm$ and reduction factor $h < 0.05$) the standard deviation is significantly greater if we use the measured albedo rather than the value obtained for the ``optimal'' set of parameters.

Thus, we are confident that the values of the thermal conductivity shown in Table~\ref{sample_results} are in reasonable agreement with the ``true'' thermal conductivity of the samples. As stated above, absolute errors are hard to give; however, from the variation of the model thermal conductivities for identical samples, we guess that the relative error of our derived thermal-conductivity values should be $\lesssim 30\%$.

\begin{figure}[!htbp]
    \center
    \includegraphics[angle=90,width=1.\columnwidth]{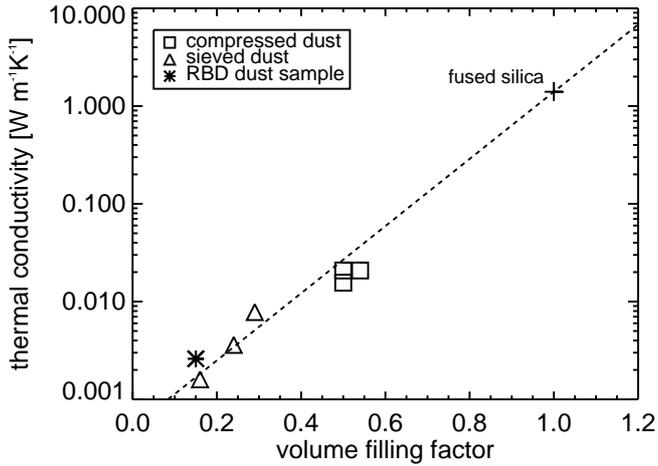}
    \caption{\label{thcond_vers_vff} Thermal conductivity of fused silica and all dust samples studied in this work plotted versus their volume filling factor. The dashed line displays the fitted exponential law $k = 0.000514 \, e^{7.91 \, \phi}$\,\WmK.}
\end{figure}

If we plot the thermal conductivity values from Table \ref{sample_results} and the value for fused silica of 1.4\,\WmK as a function of the bulk volume filling factor, we get the behavior depicted in Fig.~\ref{thcond_vers_vff}.
Together with the value of silica glass, the thermal conductivity seems to be exponentially dependent on the volume filling factor of the sample with the corresponding relation $k = 0.000514 \, e^{7.91 \, \phi}$\,\WmK. Further measurements with different materials and porosities should close the gap for higher volume filling factors and verify this exponential correlation.

\section{\label{sec:discussion}Discussion}
We developed a measuring technique to non-invasively investigate the thermal conductivity of very porous and fragile dust samples. A combination of experimental measurements and numerical simulations, precisely emulating the laboratory experiments, provides the determination of the heat conductivity of the sample.

In this work, we determined the heat conductivity of samples consisting of the same bulk material (monodisperse $\rm SiO_2$ spheres with 1.5\,$\rm \mu m$ diameter) but with different inner structures, resulting from different manufacturing techniques (RBD, sifting, compression). Our measurements and the according simulations could resolve the thermal conductivity for sample volume filling factors of $\phi~=~0.15$ to $\phi~=~0.54$, showing by trend an exponential correlation between the thermal conductivity of the sample and its porosity, with conductivity values as low as $k \sim 10^{-3}$\,\WmK~for the samples with the lowest packing density.

\begin{figure}[!t]
    \center
    \includegraphics[width=1.\columnwidth]{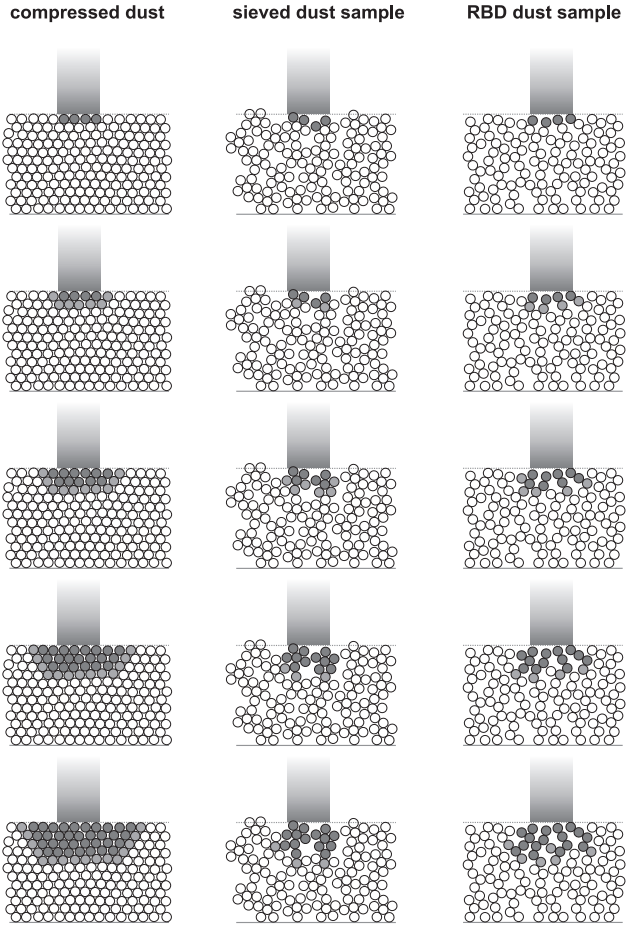}
    \caption{\label{heat-conductivity-sketch} Sketch to illustrate the stepwise heat conduction inside the three differently structured sample types. Left column: compressed dust sample; middle column: sieved dust sample; right column: dust sample produced by the RBD method.}
\end{figure}

Regarding the heat conductivity as a means of energy transport, the heat conduction is determined by the contact area between adjacent single particles of the bulk material. As the dust used in this study has a very narrow size distribution, we assume equal-sized contact areas for all contact points. Thus, the number of contacting grains (coordination number) influences the efficiency of heat transport within the material. The coordination number of a material depends on the microscopic and macroscopic homogeneity of the structure. In the case of our samples, we assume a homogenous distribution of single particles within the compressed samples and the samples made by the RBD method. In these cases, the assumption that the thermal conductivity depends directly on the coordination number and, thus, the porosity seems to be valid. For samples with a more complex inner structure, like the sieved dust samples, the correlation of thermal conductivity, porosity and coordination number is more complex. The sieved dust samples consist of a combination of porosities: the volume filling factor of the sifted dust aggregates ($\phi_{agg}$) and the packing density of the sifted aggregates inside the repository ($\phi_{aa}$). The heat conductivity within a single sieved dust aggregate differs from the heat conduction between these aggregates, due to more contact points in between the monomers inside individual sub-mm-sized aggregates. Fig.~\ref{heat-conductivity-sketch} illustrates the distinction of the heat conductivity dependent on the inner structure of the dust sample.
The sketch displays stepwise the heat conduction between adjacent monomers. Each time step comprises the heat transport through one neighboring monomer layer. By comparison, the compressed dust sample shows the most particles, which were involved in the heat transport, as they obviously have the most contact points between monomers. If we consider the sieved and the RBD dust sample, both show approximately the same amount of heated single particles, but the RBD sample slightly more. If one supposes that the thermal conductivity only depends on the porosity, one would expect the contrary: that the sieved dust sample with a lower porosity compared to the RBD sample would transfer the heat better. The sketch and the measured results of sample S3 compared to the samples RBD1 and RBD2 indicate that not the porosity but the number of contacting particles has the major influence on the thermal conductivity. From time step to time step the RBD dust sample has uniformly only a few adjacent particles which transfer the heat. In contrast, the sieved dust sample has local regions (within the sifted dust aggregates) where many contacting particles transfer the heat in one time step whereas between these aggregates only rare contact points exist, resulting in a less efficient heat conduction compared to the RBD sample.

We performed the experiments described in Sect.~\ref{sec:results} under high-vacuum conditions so that the thermal conductivity of the residual gas was negligible. It is not a priori clear whether this is also true for solar-nebula conditions. The thermal conductivity of a diatomic gas with pressure $p$ and temperature $T$ is given by
\begin{equation}\label{thckg}
  k_g = \frac{5}{6}~\frac{p\,u\,l}{T},
\end{equation}
with $u$ and $l$ being the average molecular velocity and the inter-particle distance within the dust sample, respectively. For our dust samples we expect $l=1 \ldots 10\,\rm \mu m$. To estimate the contribution of the gas to the thermal conductivity we use the solar nebula models by \citet{Weidenschilling:1977b} (MMSN) and \citet{Desch:2007}. Following the values of \citet{ZsomEtal:2010} for the midplane density of a protoplanetary disk at 1~AU of $\rho_{MMSN} = 1.4 \times 10^{-9} \, \rm g \, cm^{-3}$ and $\rho_{Desch} = 2.7 \times 10^{-8} \, \rm g \,cm^{-3}$ for $T = 200 \, \rm K$, we get for the thermal conductivity of the gas $k_{g,MSSN} = 7.0 \times 10^{-6}$\,\WmK~and $k_{g,Desch} = 1.3 \times 10^{-4}\,$\,\WmK~for $l=1 \, \rm \mu m$ and $k_{g,MSSN} = 7.0 \times 10^{-5}$\,\WmK~and $k_{g,Desch} = 1.3 \times 10^{-3}$\,\WmK~ for $l = 10 \, \rm \mu m$, respectively.
To evaluate the influence of the gas conductivity on the total thermal conductivity, we applied the model by \citet{Russell:1935} for which \citet{ChengVachon:1970} found a good agreement with measurements. For our compact samples, the increase of the total conductivity due to the gas filled pores is less than 4\% for the whole range of solar nebula gas densities. The influence of the gas on the total conductivity becomes slightly more important for those samples with high porosity and low thermal conductivity. The thermal conductivity of sample S3 is increased by 75\% if we assume a thermal conductivity of the solar nebula gas of $k_{g,Desch} = 1.3 \times 10^{-3}$\,\WmK; all other gas densities result in a thermal conductivity increase of less than 10\%. Thus, we can conclude that the influence of the gas on the thermal conductivity is negligibly small for all cases but the highest gas densities and the lowest volume filling factors, and even then only a small correction is required. The role of gas may be more important in the case of porous ice. For instance, \citet{Bar-NunLaufer:2003} measured for amorphous ``cometary'' water ice samples an increase of the thermal conductivity by a factor of 3-4 due to the effect of internal trapped gas.

The measurement of the thermal conductivity and its dependence on the porosity has significant implications for the thermal evolution of meteorite parent bodies in the early Solar System. Presently, it seems clear that meteorite parent bodies accreted within less than a few Myrs after the first solid objects -- Ca,Al rich inclusions -- formed (e.g. \citet{Trieloff:2009,Scott:2006}). Hence, short-lived nuclei, mainly $\rm ^{26}Al$ and $\rm ^{60}Fe$, were effective energy sources to heat early formed planetesimals. Classical thermal models \citep{MiyamotoEtal:1981,GhoshMcSween:1998} apply thermal conductivities measured on meteorites, which are already solidified rocks with only small degrees of porosity. However, meteorite parent bodies started with likely higher porosity, so heating effects can be expected for comparatively small planetesimals. This potentially implies significant heating effects of smaller bodies during their growth to asteroid sizes, which would imply drastically different thermal histories and has to be taken into account when modeling or describing the history of planetary building blocks in the early Solar System. Hence, our measured thermal conductivities should be implemented into models that describe the early heating of meteorite parent bodies, i.e. asteroids. For comets, which retained highly volatile species like CO, $\rm CO_2$, and $\rm H_2O$-ice particles, as recently discovered in Comet Hartley-2 (A'Hearn et al., in preparation), heating effects were obviously very minor, so that they retained their originally porous structure that likely was also a property of meteorite parent bodies before heating by short-lived nuclide decay energy. The pristinity of comets in terms of porous structure and low thermal conductivity is demonstrated by the agreement of the thermal conductivity values derived by \citet{GroussinEtal:2007} for comet 9P/Tempel~1 ($k < 1.8 \times 10^{-3} \ldots 3.6 \times 10^{-2}$\,\WmK) and the thermal conductivity of our low-density samples ($k = 1.6 \times 10^{-3} \ldots 2.1 \times 10^{-2}$\,\WmK).

\subsection*{Acknowledgements}
We thank Andreas H\"{o}pe from the PTB Braunschweig for providing us with the results for albedo measurements of compressed dust samples. This project was funded by the the Deutsche Forschungsgemeinschaft (DFG) within the Forschergruppe FOR 759 ``The Formation of Planets: The Critical First Growth Phase'' under grant Bl 298/8-1. MT acknowledges support by DFG SPP1385 and YS was supported by DFG under grant BI 298/9-1.

\bibliographystyle{model2-names}
\bibliography{literatur}

%% not want to use model2-names.bst.



%% \bibitem must have one of the following forms:
%%   \bibitem[Jones et al.(1990)]{key}...
%%   \bibitem[Jones et al.(1990)Jones, Baker, and Williams]{key}...
%%   \bibitem[Jones et al., 1990]{key}...
%%   \bibitem[\protect\citeauthoryear{Jones, Baker, and Williams}{Jones
%%       et al.}{1990}]{key}...
%%   \bibitem[\protect\citeauthoryear{Jones et al.}{1990}]{key}...
%%   \bibitem[\protect\astroncite{Jones et al.}{1990}]{key}...
%%   \bibitem[\protect\citename{Jones et al., }1990]{key}...
%%   \harvarditem[Jones et al.]{Jones, Baker, and Williams}{1990}{key}...
%%


% \end{thebibliography}

\end{document}